\documentclass[aps,prl,twocolumn,groupedaddress,showpacs]{revtex4}

\begin{document}


\title{General Relativity from the three dimensional linear group}


\author{Niall \'O Murchadha}
\email[niall@ucc.ie             ]
\affiliation{Physics Department, University College, Cork, Ireland.}


\date{\today}

\begin{abstract}
This letter describes a novel derivation of general relativity
by considering the (non)self-consistency of theories whose Hamiltonians
are constraints.
The constraints, from Hamilton's
equations, generate the evolution, while the evolution, in turn, must
preserve the constraints.
This closure requirement can be used as
 a selection mechanism for general relativity
starting from a very simple set of assumptions.
The configuration space is chosen to be a family of
$3 \times 3$ positive definite symmetric matrices on some bare 3-manifold.
A general Hamiltonian is constructed on this space of matrices which consists
of a single constraint per space
point. It is  assumed that this constraint looks like an energy balance
relationship. It will be the sum of a `kinetic' term which is
quadratic and undifferentiated in the momenta, and a `potential' term, which is
any function of the configuration variables. Further, the constraint must be a
scalar under the linear group, the natural
symmetry group of the configuration space.
 This inexorably leads to the ADM
Hamiltonian for general relativity. Both the space of
Riemannian geometries (Wheeler's superspace), and spacetime are emergent quantities 
in this analysis.
\end{abstract}


\maketitle

\section{}
One can write general relativity in Hamiltonian form by choosing the configuration
space as Wheeler's superspace, the space of 3-dimensional Riemannian metrics, 
$g_{ij}$, modulo diffeomorphisms. The Hamiltonian is \cite{adm}
\begin{equation}
{\cal H} = \int\left({N \over \sqrt{g}}
\left[\pi^{ab}\pi_{ab} -{1 \over 2}({\rm tr}\pi)^2 -gR\right] -2
 N_i\left[\pi^{ij}_{;j}\right]\right)d^3x,\label{h}
\end{equation}
where $\pi^{ij}$, a symmetric tensor density, is the momentum conjugate to $g_{ij}$,
$R$ is the scalar curvature of $g_{ij}$, $g$ is the determinant of $g_{ij}$,
 and $N$ and $N_i$, the lapse and shift, are Lagrange multipliers multiplying the 
 Hamiltonian and momentum constraints
 \begin{equation}
 H = g^{-{1 \over 2}}\left[\pi^{ab}\pi_{ab} -
{1 \over 2}({\rm tr}\pi)^2\right] -g^{1 \over 2}R = 0,\label{ham}
\end{equation}
\begin{equation}
H^i = -2\pi^{ij}_{;j} = 0. \label{mom}
\end{equation}
The evolution equations preserve the constraints. One gets
\begin{equation}
\partial_tH = {1 \over N}\left(N^2H^i\right)_{;i} +{N{\rm tr}\pi \over 2}H
+{\cal L}_{N^a}H, \label{dH}
\end{equation}
\begin{equation}
\partial_tH_j = {1 \over N}\left(N^2H\right)_{;j}
+ {\cal L}_{N^a}H_j, \label{dHi}
\end{equation}
where ${\cal L}$ is the Lie derivative in the direction of the shift.

In 1976, Hojman, Kucha\v r, and Teitelboim, \cite{hkt}, showed that if one considered 
a constrained Hamiltonian on superspace, whose Poisson bracket algebra mimicked
Eqs.(\ref{dH},\ref{dHi}), then one was led to the ADM Hamiltonian, Eq.(\ref{h}).
They imposed these specific closure requirements because they wished to guarantee
that they would recover spacetime and 4-diffeomorphism invariance. 

In a recent article,
\cite{bfom}, we showed that this was redundant. We chose superspace as the
configuration space, left the momentum constraint as it was, but considered a
(fairly) general Hamiltonian constraint. We assumed that the constraint was the sum of
a kinetic term which was quadratic in the undifferentiated momenta and a potential 
term which was some general function of the metric. The requirement that the
constraint algebra closed, without insisting on any particular form, was enough to
recover the ADM Hamiltonian.

One can do even better. In \cite{om} I considered the ADM Hamiltonian without the
momentum constraint. The Hamiltonian constraint is the only constraint and the
evolution equation for it is just Eq.(\ref{dH}) without the Lie derivative term.
The right hand side no longer weakly vanishes, we have a secondary constraint
\begin{equation}
{1 \over N}\left(N^2H^i\right)_{;i} \approx 0 \Rightarrow H^i \approx 0.
\end{equation}
Following Dirac, \cite{d}, we know we have to add this back to the Hamiltonian,
with a Lagrange multiplier, and thus we recover the standard Hamiltonian.

I wish to push this further. The obvious thing to do is to merge these two results
and consider a single scalar constrained Hamiltonian on superspace. I
intend here to go much further. I will abandon superspace, and replace it with the 
space of matrices with only the linear group as the symmetry. The requirement
that the constraint algebra closes will force me back to 3-covariance, the ADM 
Hamitonian, and, eventually, to spacetime.

I choose as my configuration space a bare 3-dimensional manifold on which I
put a family of $3 \times 3$ positive-definite symmetric matrices. I will write
them as $M_{ab}$. I can compute at each space point an inverse matrix which I
denote by $M^{ab}$. These can be distinguished because they behave differently
under rigid rotations. The momenta are in the dual of the tangent space to the
configuration space so I write them as $P^{ab}$
because they should transform just like the $M^{ab}$'s, multiplied by the
determinant of the transformation. I intend to construct a scalar Hamiltonian. To do this
I only need ensure that the `up' indices match the `down' indices.

 I further assume
that the `kinetic' part, $K$, of the Hamiltonian is quadratic in the undifferentiated
momenta. There are only two ways we can sum the momenta. Therefore we have
\begin{equation}
K = AM_{ab}M_{cd}P^{ac}P^{bd} - {B \over 2}\left[M_{ab}P^{ab}\right]^2, \label{K}
\end{equation}
where $A$ and $B$ are, as yet, arbitrary constants. I assume that the Hamiltonian also
has a potential term $Q$ which is some function on the space of matrices. 
My Hamiltonian, therefore, has the form
\begin{eqnarray}
{\cal H} = \int{N \over \sqrt{{\rm det}M}}
( AM_{ab}M_{cd}&P^{ac}P^{bd} - {B \over 2}\left[M_{ab}P^{ab}\right]^2 \cr
&- \left[{\rm det}M\right]Q)d^3 x, \label{Hm}
\end{eqnarray}
where $N$ is a Lagrange multiplier.

I will now consider the possible choices for the function $Q$.
The simplest term that $Q$ could have is an undifferentiated scalar, $CM^{ab}M_{ab}$.
This works well, it allows the final theory to have a cosmological constant. I will
put it aside for the time being and seek a more complicated expression. The next
simplest term is a term that is linear in the first derivative of $M$, $M_{ab,c}$.
Such a term cannot be used because the upstairs indices come in pairs so cannot
be used to sum over the three downstairs indices. Therefore the first nontrivial 
term in the potential must be at the next order.

This term will consist of terms which are linear in the second derivatives of the
matrices. It turns out that there are only two such
\begin{equation}
Q = CM^{ab}M^{cd}M_{ac,bd} - DM^{ab}M^{cd}M_{ab,cd} \label{Q}
\end{equation}
From a dimensional argument we should also include terms which are quadratic in
the first derivative such as
\begin{equation}
Q' = M^{ab}M^{cd}M^{ef}M_{ab,c}M_{de,f}.
\end{equation}
There are five such terms.
Therefore the Hamiltonian to this order will have nine constants, two in the
kinetic term, seven in the potential term. The logic now is quite straightforward;
one works out Hamilton's equations, applies them to the constraint and sees whether
there is any possibility that one gets no, or especially simple, secondary 
constraints for any particular choice of the constants so that the final constraint
algebra closes.

It turns out that one can simplify this problem hugely by linearizing everything.
This means assuming that the matrices $M_{ab}$ can be written as $M_{ab} = I +
\epsilon m_{ab}$, where $I$ is the identity matrix, $\epsilon$ is some small parameter
and $m_{ab}$ is of order unity. The momentum $P^{ab} = \sqrt{\epsilon}p^{ab}$.
I do not assume that $N$ is small. In turn, this means that we can drop the terms
quadratic in the first derivative and discard many of the terms arising in Hamilton's
equations.

Therefore I assume the following Hamiltonian
\begin{eqnarray}
&{\cal H} = \int{N \over \sqrt{{\rm det}M}}
( AM_{ab}M_{cd}P^{ac}P^{bd} - {B \over 2}\left[M_{ab}P^{ab}\right]^2 -\cr
 &{\rm det}M[CM^{ab}M^{cd}M_{ac,bd} - DM^{ab}M^{cd}M_{ab,cd}])d^3 x, \label{Hma}
\end{eqnarray}
This has four constants, $(A, B , C, D)$. We can set two of them to, say, 1 by an
overall scaling of the Hamiltonian and absorbing a constant into the definition
of the momentum. One should not do this immediately because the ones one picks
may turn out to vanish. 

Now I work out the evolution equations (to leading order)
 and calculate the time derivative of the constraint.
 Hamilton's equations are:
 \begin{eqnarray}
 {\partial M_{ab} \over \partial t} &=& {2 AN \over \sqrt{M}}M_{ac}M_{bd}P^{cd}
 - {BN \over \sqrt{M}}M_{ab}M_{cd}P^{cd}, \cr
 {\partial P^{ab} \over \partial t} &=& C\sqrt{M} N_{,ef}M^{ae}M_{bf} -
 \sqrt{M} N_{,ef}M^{ab}M^{ef},
 \end{eqnarray}
 plus higher order terms. Computing the evolution equation for the constraint
 gives
 \begin{eqnarray}
 {\partial H \over \partial t} = -{2AC \over N}\left(N^2P^{ab}_{,b}\right)_{,a} +
 \hskip 3cm\cr
 {BC + 2AD - 3BD \over N}M^{ef}\left(N^2\left[M_{ab}P^{ab}\right]_{,e}\right)_{,f}.
 \label{dH1}
 \end{eqnarray}
 
 I wish to consider the first term on the right hand side of Eq.(\ref{dH1}). We
 have three choices: 1) $A =0$: This is deeply boring, we end up with no dynamics,
 all the terms eventually vanish. 2) $C = 0$: This leads to a well-known and viable
 theory, sometimes called `strong gravity' \cite{i}. In this theory, we have a nontrivial
 kinetic term in the Hamiltonian, but the potential term vanishes. In addition, we 
 have no need for the momentum constraint. 3) $A \ne 0, C \ne 0$: from rescaling,
 we can set $A = C = 1$. This is the most interesting possibility, it
 forces the appearance of a secondary constraint 
 \begin{equation}
 \left(N^2P^{ab}_{,b}\right)_{,a}= 0 \Rightarrow P^{ab}_{,b} = 0 \label{mc}
 \end{equation}
 (the linearized momentum constraint!)
 on the theory.
  
  This must now be added back to the Hamiltonian, I write this in the form $-2N_a
  P^{ab}_{,b}$, where $N_a$ is the new Lagrange multiplier. The altered Hamilton's
  equations have to be derived and the time derivative of the constraints calculated
  anew. The evolution equation for the new constraint is especially interesting.
  It reads
  \begin{equation}
  {\partial \over \partial t}P^{ab}_{,b} = (C - D)M^{ag}M^{ef}N_{,egf}
  , \label{m1}
  \end{equation}
  where we have dropped a term, $+ M^{ef}P^{ab}_{,be}N_f$, which vanishes when
  the constraints hold.
  The right hand side of Eq.(\ref{m1}) must vanish (otherwise we get a neverending
  sequence of constraints). Therefore we require
  $C = D$. 

  The second term on the right hand side of Eq.(\ref{dH1}) has still to be dealt
   with. If $BC + 2AD - 3BD \ne 0$ then we get
\begin{equation}
M^{ef}\left(N^2\left[M_{ab}P^{ab}\right]_{,e}\right)_{,f} = 0 \Rightarrow
\left[M_{ab}P^{ab}\right] = {\rm constant}.
\end{equation}
This leads to two different theories. One is GR in the constant mean curvature gauge,
the other is a conformally invariant theory which we call {\it conformal gravity}
\cite{b}.

In this letter, I am much more interested in the other choice, i.e., where
$BC + 2AD -3BD = 0$. Using $A = C = D = 1$, we immediately get $B = 1$. Hence, my
selfconsistent Hamiltonian becomes
\begin{eqnarray}
&{\cal H} = \int\left[{N \over \sqrt{{\rm det}M}}
( M_{ab}M_{cd}P^{ac}P^{bd} - {1 \over 2}\left[M_{ab}P^{ab}\right]^2\right. -\cr
 &{\rm det}M[M^{ab}M^{cd}M_{ac,bd} - M^{ab}M^{cd}M_{ab,cd}])\cr
 &\left.-2N_aP^{ab}_{,b}\right]d^3 x, \label{Hma1}
\end{eqnarray}
This is nothing more than the linearized ADM Hamiltonian!

Now I check whether the constraint algebra still closes if the higher order terms
in Hamilton's equations are included. Not surprisingly, it does not work. Next, I 
add back in the five terms which are quadratic in the first derivatives with 
arbitrary coefficients and discover that there is a unique choice of coefficients
which makes everything close. I need to add
\begin{eqnarray}
&\Delta&H = M^{ab}M^{cd}M^{ef}(M_{de.c}M_{ab,f} - M_{de,c}M_{af,b}\cr &-&{1 \over 4}
M_{ab,e}M_{cd,f} - {1 \over 2}M_{de,a}M_{bf,c} +{3 \over 4}M_{de,a}M_{cf,b})
\end{eqnarray}
and
\begin{equation}
\Delta H^a = {1 \over 2}M^{ad}(2M_{bd,c} - M_{bc,d})P^{bc},
\end{equation}
and I get a fully selfconsistent constrained Hamiltonian on the space of matrices.
No other combination works.

Each individual term in the Hamiltonian is only invariant under the linear group. 
However, the total Hamiltonian is invariant under a much larger group, the group of
local 3-diffeomorphisms. This is because the combination of terms that makes up the
potential term is scalar curvature, written in terms of first and second  derivatives
of the metric. Therefore I can promote the matrices to the space of
positive-definite symmetric tensors. I can identify the 3-matrices with a 3-metric
and recognise that the configuration space is isomorphic to superspace.

This theory has arbitrary Lagrange multipliers, $N$ and $N_a$. Therefore the evolution
from a given set of initial data, viewed as a curve in superspace, is underdetermined.
This is Wheeler's `many fingured time'. Each solution curve on superspace can be used
to generate a pseudo-Riemannian manifold by the rule $g^{00} = -1/N^2, g_{0a} = N_a$.
One finds that each 4-manifold constructed in this fashion satisfies the Einstein
equations and that the solution curves arising from a given set of initial data can be
mapped into one another by 4-diffeomorphisms. Therefore, starting from a constrained
Hamiltonian on the space of matrices, there is a straight path that leads to three 
dimensional Riemannian space, four dimensional spacetime, the Lorentz group, and the
Einstein equations.

One obvious question that may be asked is why this choice of a constrained 
Hamiltonian? The naive answer is that it works. A somewhat more sophisticated answer 
may be found by making a Legandre transformation from the Hamiltonian to a
Lagrangian. Further, a parametrised action can be constructed which will have the
classic square root form of the Jacobi action (see \cite{La}). The major difference
is that the square root is inside rather than outside the integral sign.

The parametrised Jacobi action is of the form $I = \sqrt{Q}\sqrt{K'}$, where $Q$ is the
potential energy and $K'$ is the kinetic energy, a function quadratic in the velocities.
Therefor I consider a general
parametrised action of the `local square root' form, i.e., \cite{om}
\begin{equation}
{\cal L} = \int\sqrt{{\rm det} M}\sqrt{Q}
\sqrt{\left[\alpha M^{ac}M^{bd} -
\beta M^{ab}M^{cd}
\right]{\partial M_{ab} \over \partial t}{\partial M_{cd} \over \partial t}}\label{L1}
\end{equation}
where $\alpha$ and $\beta$ are again arbitrary constants.
We can work out the momentum via
\begin{equation}
P^{ab} = {\delta L \over \delta {\partial M_{ab} \over \partial t}}=
{\sqrt{{\rm det} M}\sqrt{Q} \over\sqrt{K'}}
\left[\alpha M^{ac}M^{bd} -
\beta M^{ab}M^{cd}\right]{\partial M_{cd} \over \partial t},
\end{equation}
with
\begin{equation} K' =
\left[\alpha M^{ac}M^{bd} -
\beta M^{ab}M^{cd}
\right]{\partial M_{ab} \over \partial t}{\partial M_{cd} \over \partial t}.
\end{equation}
The momentum is homogeneous of degree zero in the velocity,
which is the key property of a `square root' action. This guarantees that
the momenta are not independent. If I pick
\begin{equation}
\alpha = {1 \over A}; \hskip 0.5cm \beta = {B \over 3AB - 2A^2},
\end{equation}
one can show just by direct substitution that
\begin{equation}
 AM_{ab}M_{cd}P^{ac}P^{bd} - {B \over 2}\left[M_{ab}P^{ab}\right]^2 -
 \left[{\rm det} M\right]Q = 0.
 \end{equation}
 This is a primary constraint, which is identical to the constraint which generated
 the Hamiltonian, Eq.(\ref{Hm}).
  It is easy to show that the Euler-Lagrange equations
 of the Lagrangian, Eq.(\ref{L1}), exactly agree with Hamilton's equations of the
 Hamiltonian, Eq.(\ref{Hm}). If we make the `good' choice of the constants
 $A = B = 1$, we get $\alpha = \beta = 1$. This is the standard switch that occurs
 in going from the ADM Lagrangian to the ADM Hamiltonian where the kinetic energy
 goes from $K^{ab}K_{ab} -
({\rm tr}K)^2$ to $\pi^{ab}\pi_{ab} -
{1 \over 2}({\rm tr}\pi)^2$.

A more sophisticated (but, I think, fairly dishonest) argument could be made as 
follows: If one has a solution to the Maxwell equations on a closed manifold
without boundary, the total charge must vanish, from just the integral of Poisson's
equation. Something similar might be expected to happen in gravity. Therefore the
total gravitational energy should vanish which might give an integral constraint
of the form $\int(K - Q') = 0$, where $K$ is the kinetic energy, quadratic in the
momenta, and $Q'$ is the negative of the potential energy (some function of the
configuration variables). This means we can write $K - Q' = {\rm div} V$ where $V$ is
some vector. If this vector were a function of the configuration variables only,
and it would have to be if we were not to have derivatives of the momenta, then
we can absorb it into the $Q$ via $Q  = Q' + {\rm div} V$. This promotes a global 
constraint to a local one of the desired form.

Caveats: There are a number of (minor) warnings that should issued. As mentioned 
earlier, an undifferentiated term can be added to the potential which will just
mimic a cosmological constant, there is no restriction on either magnitude or sign.
I started with a three dimensional space and ended with four dimensional spacetime. 
This procedure works in any number of dimensions, all that changes is the value of the 
coefficient of the $(M_{ab}P^{ab})^2$ term. In $n$ space dimensions it becomes 
$-1/(n-1)$.
Finally, this approach does not pick up the signature of spacetime. At one point I
used a scaling argument to set $A = 1, C = 1$. I could just as easily have set
$A = 1, C = -1$. This will generate Euclidean gravity.

The approach advocated here can be considered as an enforced Yang-Mills construction.
In Yang-Mills one is given a global symmetry, deliberately promotes it to a local symmetry
and evaluates the consequences. Here the outcome is the same, a global symmetry becomes a 
local symmetry, but I had no choice in the matter. The requirement that the constrained 
Hamiltonian be selfconsistent forced the localization on me.

This is certainly not the `best' result of this kind. Following \cite{bfom}, I expect that 
if one added a scalar and/or vector(s) to the initial configuration space, the scalar
wave equation, Maxwell theory, and Yang-Mills theory should emerge. One needs to look
at the case where higher derivative terms appear in the potential. I have not done
this, but based on what happened in \cite{bfom}, I would be surprised if anything
new appeared. More interestingly, one could relax the condition that the momenta
be quadratic and undifferentiated. It may be that something new emerges. Finally, one
could drop the symmetry condition on the matrices. It may be that then one is led to
something akin to the Cartan theory of gravity.



\end{document}